\documentclass[sigconf,9pt]{acmart}

\usepackage{amsmath,amssymb,amsfonts}
\usepackage{graphicx}
\usepackage{textcomp}
\usepackage{amsmath}
\usepackage{booktabs} 
\usepackage{multirow}
\usepackage[most]{tcolorbox}      
\tcbuselibrary{listings,skins,breakable}
\usepackage{listings}  
\usepackage{hyperref}       
\usepackage{url}  
\usepackage{setspace}
\usepackage{booktabs}       
\usepackage{amsfonts}       
\usepackage{nicefrac}       
\usepackage{microtype}      
\usepackage{xcolor}         
\usepackage{algorithm}
\usepackage{algpseudocode}
\usepackage{amsmath}
\usepackage{tikz}
\newcommand*\circled[1]{\tikz[baseline=(char.base)]{
            \node[shape=circle,draw,inner sep=2pt] (char) {#1};}}
\usepackage{orcidlink}
\usepackage{comment}
\usepackage{amssymb}
\usepackage{braket}
\usepackage{xcolor}
\usepackage[font=scriptsize,skip=0pt]{caption}
\usepackage{cleveref}
\usepackage{setspace}
\def\BibTeX{{\rm B\kern-.05em{\sc i\kern-.025em b}\kern-.08em
    T\kern-.1667em\lower.7ex\hbox{E}\kern-.125emX}}

\usepackage{enumitem}
\usepackage{hyperref}
\setlist[itemize]{leftmargin=*}%
\setlist[enumerate]{leftmargin=*}%

\usepackage[most]{tcolorbox}
\usepackage{hyperref}
\usepackage{cleveref}

\newtcolorbox[auto counter, number within=subsection]{mybox}[2][]{%
  enhanced,
  colback=black!3!white,
  colframe=black!90!black,
  breakable,
  left=0.5pt, right=0.5pt, top=0.5pt, bottom=0.5pt,
  boxsep=0.5pt, boxrule=0.5pt,
  title=Box~\thetcbcounter\ -- #2,
  label type=mybox,
  #1
}

\crefname{mybox}{Box}{Boxes}
\Crefname{mybox}{Box}{Boxes}

\newcommand{\rpoint}[1]{\scalebox{0.8}{\circled{{\fontfamily{pcr}\selectfont\footnotesize #1}}}}

\copyrightyear{2026}
\acmYear{2026}
\setcopyright{cc}
\setcctype{by}
\acmConference[DAC '26]{63rd ACM/IEEE Design Automation Conference}{July 26--29, 2026}{Long Beach, CA, USA}
\acmBooktitle{63rd ACM/IEEE Design Automation Conference (DAC '26), July 26--29, 2026, Long Beach, CA, USA}
\acmDOI{10.1145/3770743.3807703}
\acmISBN{979-8-4007-2254-7/2026/07}

\begin{document}
\settopmatter{printfolios=false}


\title{Late Breaking Results: Hardware-Aware Compilation Reshapes Trainability in Variational Quantum Circuits
}



%

\author{
Muhammad Kashif~\textsuperscript{1,2}, Muhammad Shafique\textsuperscript{1,2}}
\affiliation{%
  \institution{\textsuperscript{1}eBRAIN Lab, Division of Engineering, New York University (NYU) Abu Dhabi, Abu Dhabi, UAE}
  \streetaddress{PO Box 129188}
  \city{}
  \country{}
}

\affiliation{%
  \institution{\textsuperscript{2}Center for Quantum and Topological Systems, NYUAD Research Institute, NYU Abu Dhabi, Abu Dhabi, UAE}
  \city{}
  \country{}
}
\email{{muhammadkashif,  muhammad.shafique}@nyu.edu}

\begin{abstract}
\vspace{-4pt}
\begin{spacing}{0.88}
Variational quantum circuits (VQCs) are typically evaluated at the logical design level when analyzing trainability. However, execution on real quantum devices requires hardware-aware compilation (transpilation) to satisfy qubit connectivity and native gate-set constraints. 
In this paper, we examine how transpilation can alter the gradient statistics. Using parameter-shift differentiation and gradient variance estimation, we compare logical and transpiled circuits across three representative ansatz families: EfficientSU2 (dense entanglement), TTN (tree tensor network), and RealAmplitudes (linear entanglement). We observe architecture-dependent trainability shifts where densely entangling circuits exhibit pronounced gradient reshaping in shallow regimes, structured tensor-network circuits remain comparatively robust, and linear architectures show mixed behavior. Deep circuits across all families display minimal sensitivity to hardware-aware compilation. These findings demonstrate that transpilation acts as an implicit structural transformation of the optimization landscape, motivating compilation-aware analysis and co-design for VQCs.

\end{spacing}
\vspace{-4pt}
\end{abstract}

\maketitle

\begin{spacing}{0.9}
\vspace{-9pt}
\section{Introduction}
\vspace{-2pt}

Variational quantum circuits (VQCs) form the backbone of near-term quantum machine learning and hybrid quantum–classical algorithms. Their effectiveness depends on trainability, typically quantified through gradient statistics such as gradient variance~\cite{kashif_allev}. A central here challenge is the emergence of barren plateaus (BPs), where gradients vanish exponentially, negatively impacting the trainability of VQCs and eventually making the optimization increasingly difficult~\cite{McClean:2018}. Existing studies mostly analyze BPs and optimization/trainability landscapes at the logical circuit level~\cite{kashif_allev}, implicitly assuming that the circuit executed on hardware matches its abstract logical design, which is questionable in the NISQ era.

In practice, circuits must undergo hardware-aware compilation (transpilation)\footnote{we use hardware-aware compilation and transpilation interchangeably in this paper} to satisfy device-specific constraints such as qubit connectivity and native gate-set constraints. This process introduces routing overhead (additional two-qubit gates), and basis decompositions that can substantially alter circuit structure. \textit{Although transpilation is typically regarded as a resource-increasing transformation primarily affecting depth and noise accumulation, its direct impact on trainability of VQCs remains largely unexplored}.

In this paper, we isolate the structural effect of transpilation on trainability by comparing gradient variance between logical circuits and their transpiled counterparts under noiseless simulation. We evaluate three representative ansatz families with distinct entanglement structures: EfficientSU2~\cite{effSU2_ibm}, TTN (tensor-tree network), and RealAmplitudes~\cite{real_amp_ibm}. Across varying qubit counts and ansatz repitions repetitions, we observe that transpilation reshapes gradient statistics in an architecture-dependent manner. Dense circuits exhibit pronounced gradient reshaping in shallow regimes, structured tensor-network circuits remain comparatively robust, and linear architectures display mixed behavior.
%
%
\textit{Our findings demonstrate that transpilation acts as an implicit architectural perturbation of the optimization landscape rather than a simple resource transformation. Thus, our results motivate hardware-aware trainability evaluation and indicate that state-of-the-art trainability analyses conducted at the logical circuit level may require re-evaluation using transpiled hardware-constrained circuits.}
\vspace{-12pt}
\begin{figure}[h]
    \centering
    \includegraphics[width=0.87\linewidth]{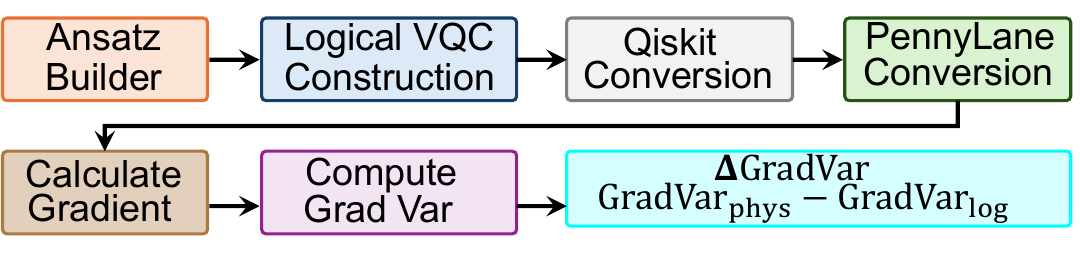}
    \vspace{-1.5pt}
    \caption{Logical-to-physical trainability evaluation pipeline. For each ansatz, the logical circuit is transpiled to a hardware-constrained physical circuit. Gradient variance is estimated using parameter-shift differentiation and multi-seed averaging to quantify transpilation-induced trainability shifts}
    \label{fig:LBR_method}
    \vspace{-20pt}
\end{figure}

\section{Our Methodology}
\vspace{-3pt}
To quantify how transpilation alters optimization behavior, we compare logical circuits with their transpiled counterparts under noiseless simulation. By evaluating gradient statistics before and after transpilation under identical experimental conditions, we directly measure transpilation-induced trainability shifts. An overview of our methodology is presented in Fig.~\ref{fig:LBR_method}.

\vspace{-9pt}
\subsection{Logical-to-Physical Circuit Construction}
\begin{figure}[h]
    \centering
    \includegraphics[width=0.95\linewidth]{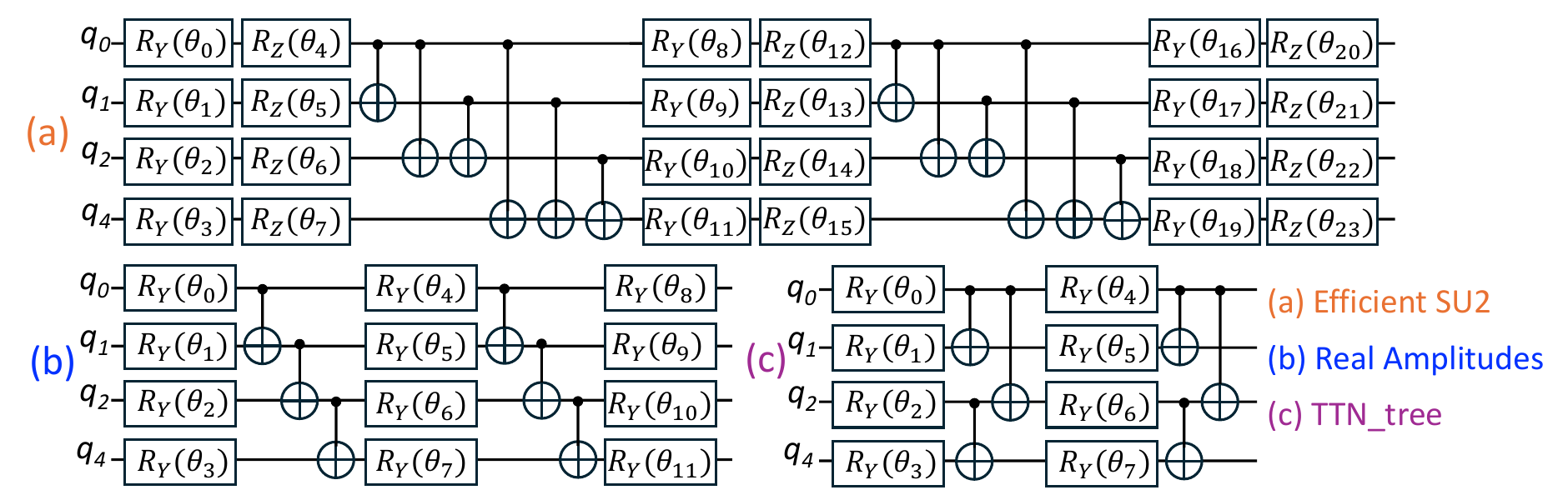}
    \caption{Examples of Ansaetze used in this paper}
    \label{fig:ansatz_LBR}
    \vspace{-12.5pt}
\end{figure}

\vspace{-2pt}
We construct VQCs using an ansatz registry comprising three representative architectures: EfficientSU2 (dense full entanglement), TTN (tree tensor network), and RealAmplitudes (linear entanglement), as shown in Fig.~\ref{fig:ansatz_LBR}. 
Each ansatz is instantiated with $n$ qubits and $L$ repetitions, producing a variational logical circuit $C_{\text{log}}(\theta)$.
\vspace{-11pt}
Logical circuits are implemented using Qiskit as hardware-agnostic designs without connectivity constraints. 
To model realistic execution, circuits are transpiled using Qiskit's transpiler with \texttt{IBMFakeBrooklynV2} as the target hardware backend. We use \texttt{opt\_level=3} for the transpiler because it enables a full range of optimization techniques \cite{transpiler}. 
Transpilation includes SWAP-based routing, native gate decomposition, and compiler optimizations, resulting in a hardware-constrained physical circuit $C_{\text{phys}}(\theta)$.
We quantify transpilation-induced structural overhead:
\vspace{-4pt}
\begin{equation}
\Delta G_{2q} = G_{2q}^{\text{phys}} - G_{2q}^{\text{log}}
\end{equation}
\vspace{-7pt}
\begin{equation}
\Delta G_{1q} = G_{1q}^{\text{phys}} - G_{1q}^{\text{log}}
\end{equation}
\vspace{-8pt}
\begin{equation}
\Delta \text{depth} = \text{depth}_{\text{phys}} - \text{depth}_{\text{log}}
\end{equation}
\vspace{-1pt}
where $G_{1q}$ and $G_{2q}$ denotes the number of single and two-qubit gates respectively. The physical depth ($\text{depth}_{\text{phys}}$) is the depth of VQC as seen by the quantum hardware, and is different from logical depth ($\text{depth}_{\text{log}}$), which is number of times a given VQC is repeated. 
\vspace{-17pt}
\subsection{Gradient-Based Trainability Evaluation}
\vspace{-3pt}
Both logical and physical circuits are converted to PennyLane for gradient evaluation. We define a scalar cost function as in Eq.~\ref{eq:cost_ftn}, and compute gradients via parameter-shift differentiation. 
\vspace{-2.5pt}
\begin{equation}\label{eq:cost_ftn}
L(\theta) = \langle Z_0 \rangle
\end{equation}
\vspace{-2.5pt}
For each circuit, parameters are sampled uniformly from $[0, 2\pi]$, and gradients w.r.t all parameters are evaluated across multiple random initializations. Trainability is evaluated using gradient variance (Eq.~\ref{eq:gradvar}), where $P$ is the number of trainable parameters:
\vspace{-5.5pt}
\begin{equation}\label{eq:gradvar}
GradVar =
\frac{1}{P}
\sum_{i=1}^{P}
\mathrm{Var}_{\theta}
\left(
\frac{\partial L}{\partial \theta_i}
\right)
\end{equation}
\vspace{-17pt}
\subsection{Transpilation-Induced Trainability Shift}
\vspace{-2pt}
To isolate the effect of transpilation, we compute difference in gradients variance using Eq.~\ref{eq:gradvardelta}. Positive values indicate gradient amplification after transpilation, while negative values indicate gradient suppression. This metric directly quantifies transpilation-induced reshaping of the optimization landscape.
\vspace{-2pt}
\begin{equation} \label{eq:gradvardelta}
\Delta \text{GradVar} =
\text{GradVar}_{\text{phys}} -
\text{GradVar}_{\text{log}}
\end{equation}
\vspace{-22pt}
\section{Results}
\vspace{-5pt}
\subsection{Structural Overhead During Transpilation}
\begin{figure}[t!]
    \centering
    \includegraphics[width=0.97\linewidth]{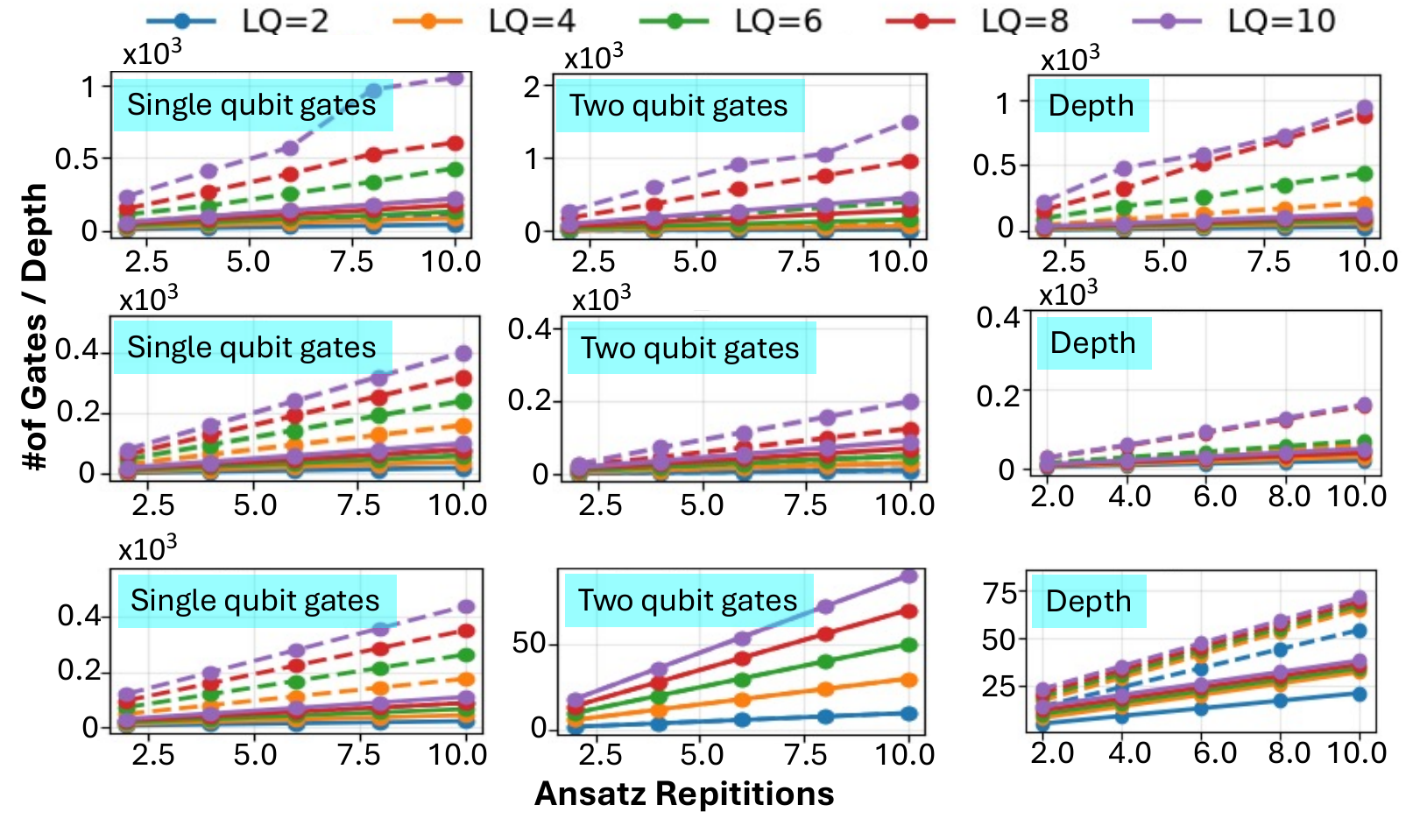}
    \vspace{-4pt}
    \caption{Logical (solid) vs physical (dashed) circuit growth under optimized transpilation (opt=3). Two-qubit gate count and depth inflate significantly with qubit number and repetitions, especially for the densely entangling EfficientSU2 ansatz. Top (EfficientSU2), middle (TTN), bottom (Real Amplitudes). LQ = Logical Qubits.}
    \label{fig:LBR_res1}
    \vspace{-17pt}
\end{figure}

\vspace{-3pt}
Fig.~\ref{fig:LBR_res1} compares logical (solid) and transpiled physical (dashed) circuits for all ansatz familiels. Across all architectures, transpilation significantly increases circuit depth and gate counts. The densely entangling EfficientSU2 ansatz exhibits the largest structural growth, TTN shows comparatively moderate overhead, and RealAmplitudes displays intermediate behavior. These results reinforces the fact that transpilation substantially alters circuit structure through routing and decomposition overhead. We next examine how this structural inflation reshapes trainability.
%
%
\vspace{-9pt}
\subsection{Trainability Landscape Reshaping}
\begin{figure}[h]
    \centering
    \includegraphics[width=0.96\linewidth]{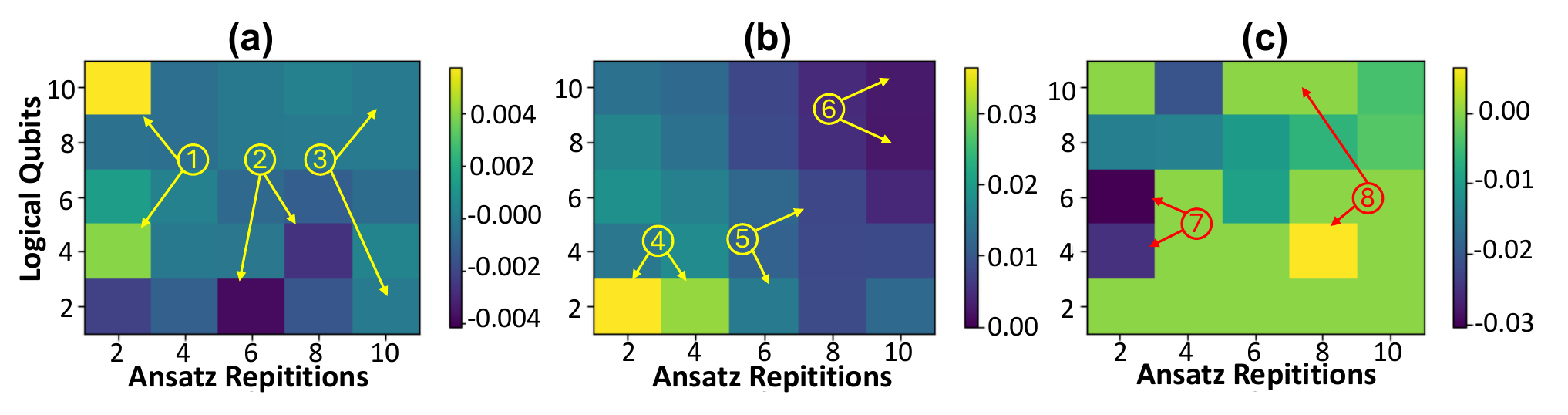}
    \vspace{-3pt}
    \caption{Trainability shift after hardware-aware compilation. Heatmaps report $\Delta \text{GradVar}$ between physical and logical circuits across repetitions and qubit counts, highlighting architecture-dependent gradient reshaping. (a) EfficientSU2, (b)TTN, (c) RealAmplitudes}
    \label{fig:LBR_results}
\end{figure}

\vspace{-2pt}
Fig.~\ref{fig:LBR_results} reports the transpilation-induced trainability shift (Eq.~\ref{eq:gradvardelta}) across different qubit counts and ansatz repetitions. Positive values indicate better trainability after transpilation, while negative values indicate reduced trainability.
%
For EfficientSU2, transpilation induces clear regime-dependent behavior. Shallow circuits exhibit positive $\Delta \text{GradVar}$ (pointer~\rpoint{1}, Fig.~\ref{fig:LBR_results}), indicating that the transpiled circuits often display higher gradient variance than their logical counterparts. Intermediate depths show localized negative squares (pointer~\rpoint{2}, Fig.~\ref{fig:LBR_results}), whereas deep circuits display near-zero shifts (pointer~\rpoint{3}, Fig.~\ref{fig:LBR_results}), suggesting limited impact once the circuit is already highly entangling.
TTN ansatz demonstrates comparatively robust behavior. Positive shifts appear primarily in shallow regimes (pointer~\rpoint{4}, Fig.~\ref{fig:LBR_results}), and $\Delta \text{GradVar}$ decays monotonically with increasing repetitions (pointer~\rpoint{5}, Fig.~\ref{fig:LBR_results}). However, no significant negative regions are observed (pointer~\rpoint{6}, Fig.~\ref{fig:LBR_results}), indicating that structured hierarchical entanglement is less sensitive to transpilation-induced perturbations in regards to gradient variance.
%
RealAmplitudes exhibits predominantly negative or near-zero $\Delta \text{GradVar}$ values, indicating that transpilation generally suppresses (pointer~\rpoint{7}, Fig.~\ref{fig:LBR_results}), or leaves gradient variance largely unchanged (pointer~\rpoint{8}, Fig.~\ref{fig:LBR_results}), depending on depth and qubit count. This highlights that transpilation effects depend strongly on the interaction between entanglement structure and routing overhead.
%
Across all three ansatz families, two consistent patterns emerge: (i) shallow circuits are most sensitive to transpilation, and (ii) deep circuits remain comparatively stable. Transpilation therefore does not uniformly degrade trainability, instead, it acts as an architecture-dependent structural transformation of the optimization landscape.
\vspace{-12pt}
\vspace{-21pt}
\section{Conclusion}
\vspace{-2pt}
We demonstrate that hardware-aware compilation is not merely a neutral resource transformation for variational quantum circuits (VQCs), but can reshape their trainability, and is  strongly architecture-dependent. Densely entangling circuits such as \textit{EfficientSU2} exhibit pronounced trainability reshaping in shallow regimes, structured \textit{TTN} architectures remain comparatively robust, and \textit{RealAmplitudes} display mixed behavior. Deep circuits across all families show less sensitivity to transpilation.
These observations suggest that hardware mapping acts as an implicit structural perturbation of the optimization landscape rather than a purely resource-based transformation. This highlights the importance of transpilation-aware trainability evaluation and hardware–algorithm co-design for near-term VQC design. 
Future work will incorporate noise models, larger circuit scales, and additional VQC architectures to better understand how transpilation interacts with VQC's optimization dynamics.
\vspace{-9pt}
\section*{Acknowledgements}
\vspace{-3pt}
This work was supported in part by the NYUAD Center for Quantum and Topological Systems (CQTS), funded by Tamkeen under the NYUAD Research Institute grant CG008.
\end{spacing}
\vspace{-5pt}
\begin{spacing}{0.6}
\bibliographystyle{ACM-Reference-Format}
\bibliography{main}
\end{spacing}

\end{document}